\documentclass[final,3p,twocolumn]{elsarticle}
\usepackage{graphicx}
\usepackage{amssymb}
\usepackage{amsmath}
\biboptions{numbers,sort&compress}
\journal{Physics Letters B}

\begin{document}
\begin{frontmatter}
\title{Bubble Structure in Magic Nuclei}
\author[a,b]{G. Saxena}
\author[a,b]{M. Kumawat}
\author[c]{M. Kaushik}
\author[b] {S. K. Jain}
\author[d]{Mamta Aggarwal\corref{cor1}}
\cortext[cor1]{corresponding authors: Mamta Aggarwal, mamta.a4@gmail.com}
\address[a]{Department of Physics, Government Women Engineering College, Ajmer-305002, India}
\address[b]{Department of Physics, School of Basic Sciences, Manipal University, Jaipur-303007, India}
\address[c]{Department of Physics, Shankara Institute of
Technology, Kukas, Jaipur-302028, India}
\address[d]{Department of Physics, University of Mumbai, Kalina Campus, Mumbai-400098, India}
\begin{abstract}
The existence of bubble nuclei identified by the central depletion in nucleonic density is studied for the conventional magic N (Z) $=$ 8, 20, 28, 40, 50, 82, 126 isotones (isotopes) and recently speculated magic N $=$ 164, 184, 228 superheavy isotones. Many new bubble nuclei are predicted in all regions. Study of density profiles, form factor, single particle levels and depletion fraction (DF) across the periodic chart reveals that the central depletion is correlated to shell structure and occurs due to unoccupancy in s-orbit (2s, 3s, 4s) and inversion of (2s, 1d) and (3s, 1h) states in nuclei upto Z $\le$ 82. Bubble effect in superheavy region is a signature of the interplay between the Coulomb and nn-interaction and depletion fraction (DF) is found to increase with Z (Coulomb repulsion) and decrease with isospin. Our results are consistent with the available data. The occupancy in s-state in $^{34}$Si increases with temperature which appears to quench the bubble effect.
\end{abstract}


\end{frontmatter}

Observation of the "Bubble" structure in atomic nuclei is a novel exotic nuclear phenomenon which is characterized by the distinct central depletions of the matter distribution~\cite{WILSON,nature,duguet,todd,grasso,khan,wang,wang1,grasso1,yao1,li,schuetrumpf,wu}. The ability to produce more exotic nuclei with advanced RIB facilities has revived the interest in the bubble nuclei which was first visualized in early nineteen forties ~\cite{WILSON}. The central depletion in the nucleonic density mainly arises due to the unoccupancy of the s-state near the Fermi surface. This causes the density at the center either to vanish or become significantly lower than the saturation density. In some cases the depopulation in the s-orbit occurs due to the inversion of s$_{1/2}$ with an another state usually located above, such as inversion of 2s$_{1/2}$ \& 1d$_{3/2}$ or 3s$_{1/2}$ \& 1h$_{11/2}$ states~\cite{khan}. On the contrary, the occurrence of bubble phenomenon in heavy and superheavy nuclei~\cite{sobi,decharge,sksingh,ikram,bender} has been attributed to Coulomb repulsion or rather an interplay between the Coulomb and nn-interaction. However the pairing correlation effects and the deformation have been observed to hinder the bubble formation. Interestingly, the bubble phenomenon is found in all the mass regions from light, medium, heavy to superheavy nuclei. \par

The occurrence of the bubble structure can be quantified by defining a depletion fraction (DF) as
\begin{equation}
DF = (\rho_{max}-\rho_{c})/\rho_{max}
\end{equation}
where $\rho$$_{max}$ and $\rho$$_{c}$ $=$ $\rho$(r $=$ 0) represent the values of the maximum  and central charge density. Since the density fluctuation is, in general, related to the quantal effects related to the filling of single-particle levels near the Fermi energy, the depletion fraction is also sensitive to the quantal effects. The s (\textit{l} $=$ 0) orbitals are the only non-zero wavefunction at the origin (r $=$ 0) with the radial distribution peaked at the center of the nucleus. However, a vacancy in the s-orbit near the Fermi level, results in a depletion of central density, whereas the non-zero \textit{l} orbitals which are suppressed in the interior of the nucleus do not contribute to nuclear density at the center. Hence the best possible bubble candidates to exhibit bubble structure are expected to have unoccupied s-orbital. This is a necessary condition for bubble effect but in addition to this, the s-orbit near the Fermi energy must be surrounded by orbitals of larger \textit{l} (the larger the better) which should be well separated in energy from its nearby single-particle states so that the dynamical correlations are weak. It is important to note that the depletion in the center associated with the vacancy in s-orbit is reinforced by the occupied orbitals whose maximum occurs at the large distances. Hence both the conditions together potentially maximize the bubble effect. Apart from the pairing and dynamical correlations, temperature has been speculated to quench the bubble structure~\cite{TAN} in agreement with one of our results presented in this letter where we have used the statistical theory (ST) of hot nuclei~\cite{MAPLB,MAPRC} for the first time to investigate the anti bubble effect of temperature. For ground state nuclei (T $=$ 0), we use Relativistic mean-field (RMF) plus state dependent Bardeen-Cooper-Schrieffer (BCS) theory~\cite{saxena,saxena1}, which has been generally found to be effective to treat such a wide range of masses that too upto the drip lines~\cite{saxena,saxena1,walecka,gambhir,yadav1,singh}. We perform a systematic study of density profiles, single particle spectra, charge form factor and depletion fraction for the isotonic and isotopic chains of magic nuclei with N(Z) $=$ 8, 20, 28, 40, 50, 82, N $=$ 126. The superheavy N $=$ 164, 184, 228 isotones are also studied. We predict many new bubble nuclei in all the mass regions. \par
\begin{figure}[htb]
\centering
\includegraphics[width=0.5\textwidth]{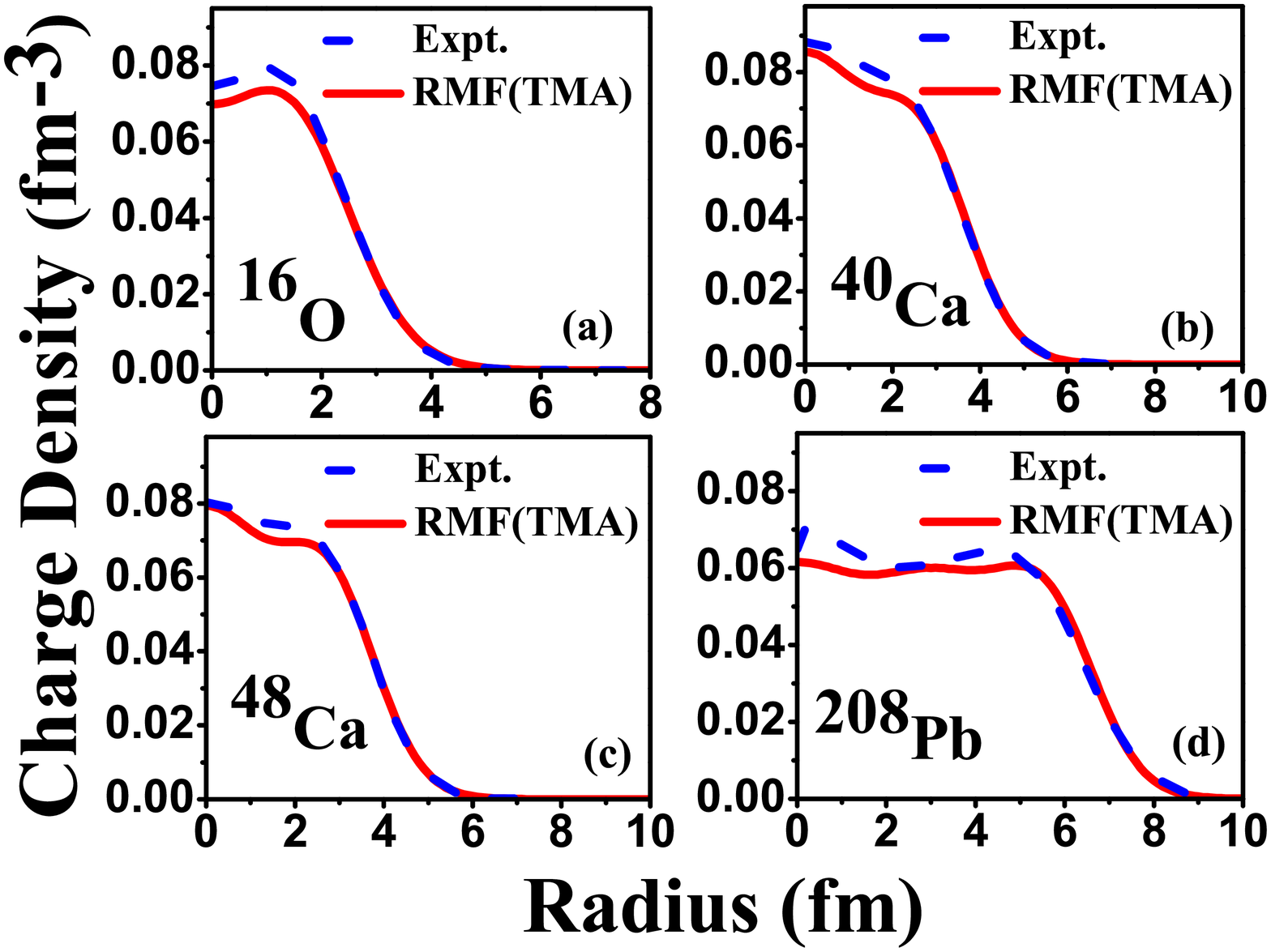}
\caption{(Colour online) Charge density of (a) $^{16}$O, (b) $^{40}$Ca, (c) $^{48}$Ca and (d) $^{208}$Pb vs. radius along with experimental data~\cite{vries}.}
\label{fig1}
\end{figure}
Recent experimental evidence for bubble in $^{34}$Si~\cite{nature} has opened a major frontier for theoretical research, that has, so far, provided a reasonable amount of information~\cite{WILSON,nature,duguet,todd,grasso,khan,wang,wang1,grasso1,yao1,li,schuetrumpf,wu} on potential bubble nuclei such as $^{22}$O, $^{34}$Si, $^{46}$Ar, $^{68}$Ar, $^{206}$Hg, and proton semi-bubble in superheavy $^{294}$Og~\cite{schuetrumpf}. The central nucleonic density in superheavy region is entirely driven by the Coulomb repulsion and is related to the symmetry energy J~\cite{schuetrumpf}. However, the single-reference (SR) energy density functional (EDF) calculations have been used to study the bubble structure in heavy nuclei. It shows that the ground-state configuration of heavy/superheavy nuclei may display bubble like structure \cite{decharge1,bender1}, as a result of a collective quantum mechanical effect, sustained by the compromise between the large repulsive Coulomb interaction and the attractive nucleon nucleon strong force. Therefore, it is speculated that the quantum shell effects, which play a major role in bubble effect in lighter nuclei, may not be predominant but may play a subtle role in central depletion of heavier systems.\par
 The inclusion of long range correlations and dynamical quadrupole shape effects have been reported to quench the bubble effect on the basis of MR-EDF ~\cite{yao,yao1} and shell-model (SM) ~\cite{grasso} calculations, but not eliminate the bubble effect~\cite{wu}. Calculations of $^{34}$Si ~\cite{duguet} with the ab initio many-body method showed that the dynamical correlations reduce the depletion factor by about 0.15 unit without erasing the bubble structure entirely. Furthermore, it is shown ~\cite{duguet} that the effect of correlations is not only to change the single-particle occupation probabilities but also the radial shape of the natural wave-functions. This effect becomes less pronounced as T increases and is expected to completely disappear at a certain critical value of T around 3-4 MeV which needs further investigation. The tensor-force and the pairing correlations have been found to have important implications in the shell evolution and the bubble structure. The existence of the proton bubble in $^{46}$Ar shows certain uncertainties. The pairing correlations quench the bubble effect in $^{46}$Ar whereas the tensor force favors it~\cite{nakada, khan,grasso,wu}. A better insight on this aspect is expected from the charge density measurements by the upcoming facilities SCRIT, RIBF ~\cite{suda,suda1} because $^{46}$Ar, may be, in principle, possible to study with RI production in near future.\par

\begin{figure}[htb]
\centering
\includegraphics[width=0.5\textwidth]{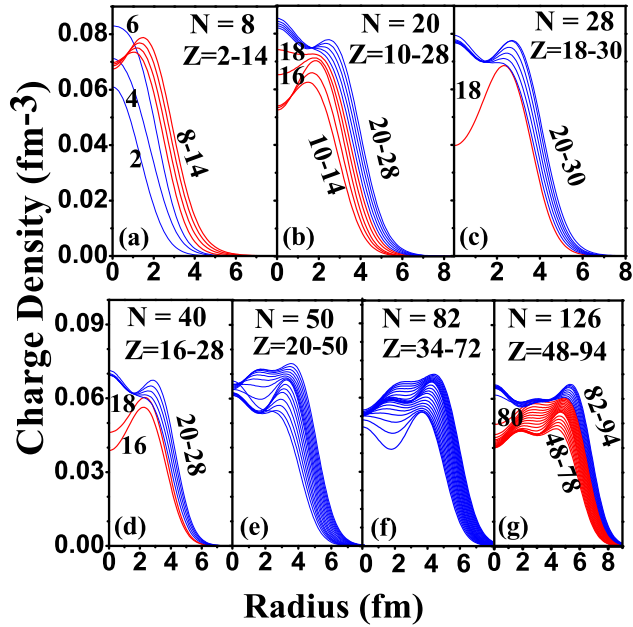}
\caption{(Colour online) Charge density vs. Radius for N $=$ 8, 20, 28, 50, 82, 126 isotones. Numbers on curves represent proton number. Red lines
denote central depletion and blue represent undepleted density.}
\label{fig2}
\end{figure}
To assess the ability of the employed RMF parameter (TMA) to reproduce the experimentally known charge densities, we have compared charge density of magic nuclei $^{16}$O, $^{40}$Ca, $^{48}$Ca and $^{208}$Pb with that of the experiments ~\cite{vries} in Fig. 1. Data of experimental~\cite{vries} density have been extracted from the Fourier-Bessel Coefficients analysis. Fig. 1. shows reasonable agreement in light, medium and heavy mass region. So, we extend our calculations to other nuclei.\par
\begin{figure}[htb]
\centering
\includegraphics[width=0.5\textwidth]{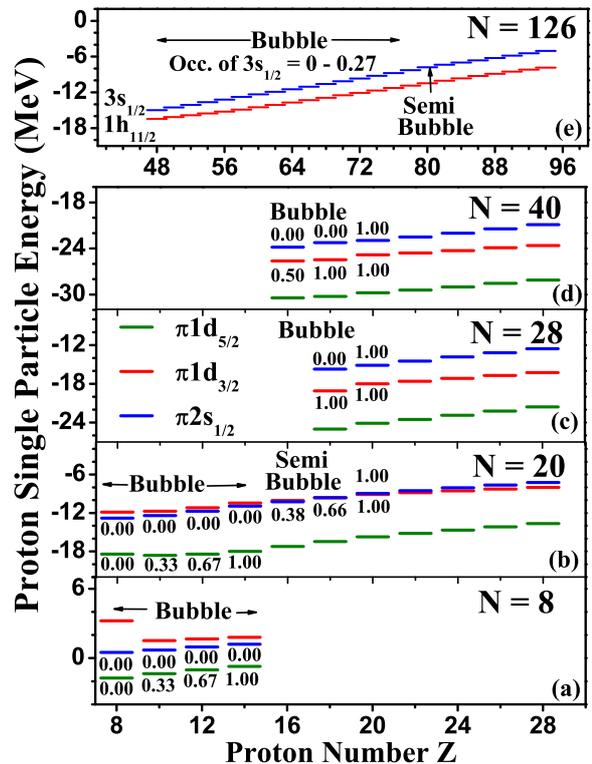}
\caption{(Colour online) Proton s.p. levels vs. Z for N $=$ 8, 20, 28, 40, 126 isotones. Occupancy (occ.) of protons is mentioned near
levels (only relevant ones are shown).}
\label{fig3}
\end{figure}
The charge densities of N $=$ 8, 20, 28, 40, 50, 82 and 126 isotones plotted in Fig. 2 show the depletion of density at the center (r $=$ 0). We find that $^{22}$Si (Z $=$ 14) in N $=$ 8 isotones, $^{30}$Ne, $^{32}$Mg and $^{34}$Si (Z $=$ 10, 12 and 14) in N $=$ 20 isotones show strong bubble structures with the central depletion. The isotones $^{46}$Ar, $^{56}$S, $^{58}$Ar with N $=$ 28 and 40 and the isotones with N $=$ 126 isotones and Z $=$ 48$-$78 show up significant central density depletion so are marked as bubble candidates. But isotones with N $=$ 50 and 82 do not show central depression indicating no bubble structure. \par
\begin{figure}[htb]
\centering
\includegraphics[width=0.5\textwidth]{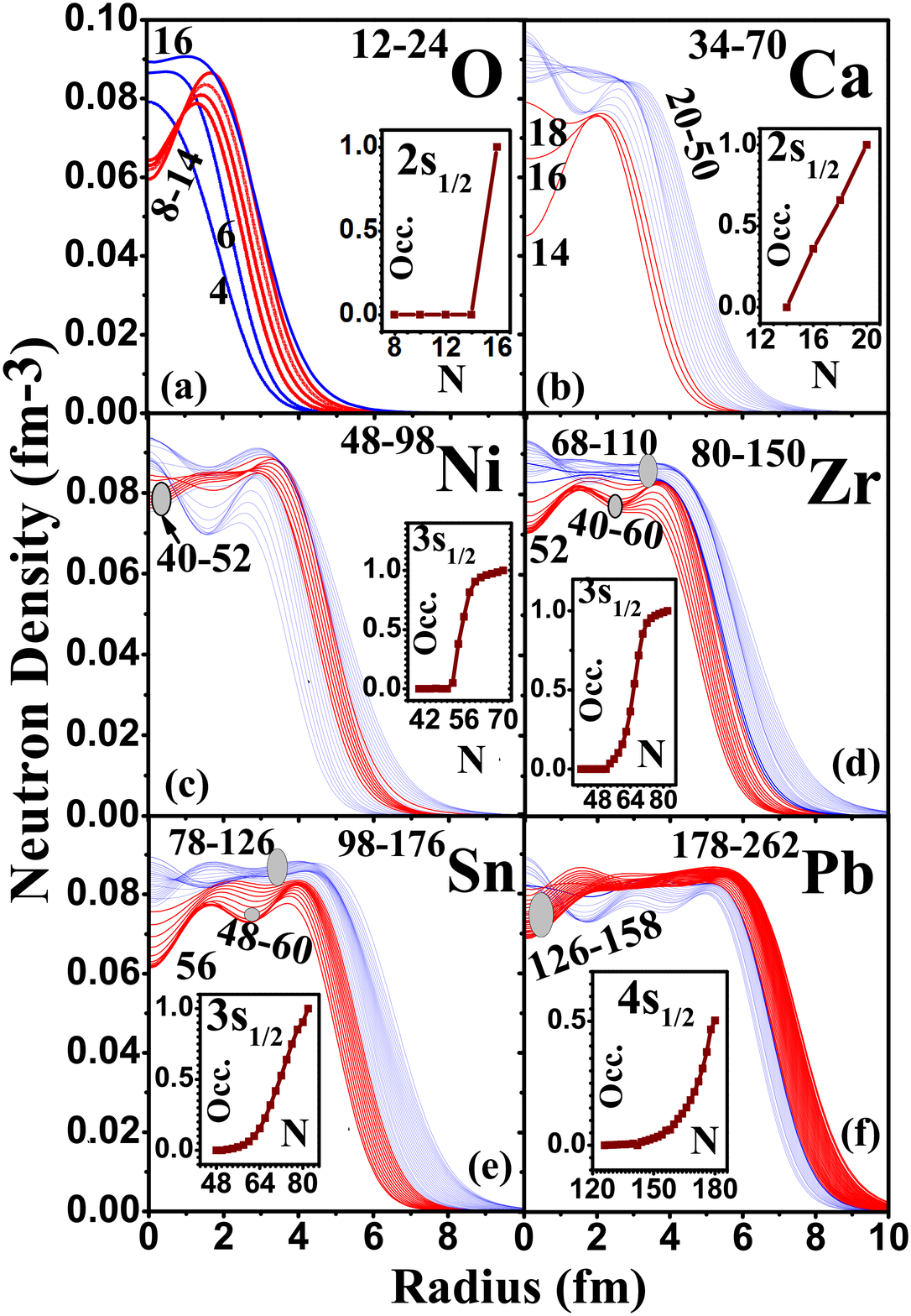}
\caption{(Colour online) Neutron density vs. radius for Z $=$ 8, 20, 28, 40,
50, 82 and 126 isotopes. Numbers on curves represent neutron number. Red lines
denote central depletion and blue represent undepleted density. Insets show occ. in s-state vs. N.}
\label{fig4}
\end{figure}
So far it is known that the bubble effect relates to the quantum shell effects with its origin in the sequence of occupied and unoccupied s.p. states, in particular, the s-orbital near Fermi level. In view of this, we investigate the occupation probability and s.p. spectra of the proton sd shell ($1d_{5/2}$, 2s$_{1/2}$, $1d_{3/2}$) for N $=$ 8, 20, 28, 40, and proton 3s$_{1/2}$ and 1h$_{11/2}$ states for N $=$ 126 in Fig. 3.  Unoccupancy of 2s$_{1/2}$ state in Z $=$ 8$-$14 (N $=$ 8) isotones (Fig. 3(a)) leads to the central density depletion seen in Fig. 2 (a). In N $=$ 20 isotones (Fig. 3(b)), the energy gap between the states ($1d_{5/2}$ and 2s$_{1/2}$) increases from Z $=$ 8 to 14 and attains a maximum value of 7 MeV at Z $=$ 14 with full occupancy in $1d_{5/2}$ state and completely unoccupied 2s$_{1/2}$ state that results in the central depletion in $^{34}$Si ( Fig. 2(b)). Large energy gap at Fermi level marks $^{34}$Si a doubly magic nucleus and a prominent proton bubble candidate as expected, in agreement with experimental~\cite{nature} and other theoretical~\cite{duguet,wang,wang1,grasso1,yao1} works. On the other hand, the energy gap between the states 2s$_{1/2}$ and $1d_{3/2}$, decreases from Z $=$ 8 to 14 and attains a small value $\approx$ 0.2 MeV at Z $=$ 14. Here both the states being too close to each other get occupied simultaneously following an inversion~\cite{todd,grasso} of the states (2s$_{1/2}$ and 1d$_{3/2}$). Consequently, 2s$_{1/2}$ remains semi-occupied and contributes partially to the depletion of central density in Z $=$ 16 and 18 resulting in the semi-bubble nuclei $^{36}$S and $^{38}$Ar. In N $=$ 28 and 40 isotones (Figs. 3(c), 3(d)), the inversion of 2s$_{1/2}$ and $1d_{3/2}$ separated by an energy gap of more than 3 MeV results in the fully occupied $1d_{3/2}$ and an unoccupied 2s$_{1/2}$ state in $^{46}$Ar, $^{58}$Ar and semi-occupied 2s$_{1/2}$ in $^{56}$S leading to density depletion seen in Fig. 2 ((c) and (d)). Here it may be noted that RMF (with TMA) shows significant central depletion in $^{46}$Ar~\cite{chu} due to inversion of 2s$_{1/2}$ and 1d$_{3/2}$ states without including the tensor force~\cite{todd,grasso,khan,wu}. Shell structure of N $=$ 50 and 82 isotones (Fig. 2 (e), (f)) results in the absence of s$_{1/2}$ state near the Fermi level and shows no bubble character. The inversion of proton states (3s$_{1/2}$ and 1h$_{11/2}$) similar to $^{206}$Hg~\cite{wang2} results in the unoccupied 3s$_{1/2}$ state in all the N $=$ 126 (Z $=$ 48$-$78) isotones indicating bubble structure in the complete chain (seen in Fig. 2(g)). \par
\begin{figure}[htb]
\centering
\includegraphics[width=0.5\textwidth]{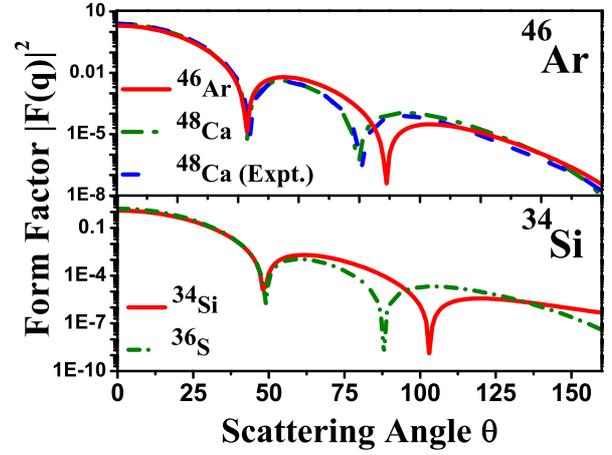}
\caption{(Colour online) Nuclear charge form factors with scattering angle for (a) $^{48}$Ca and $^{46}$Ar (b) $^{36}$S and $^{34}$Si.
Experimental data for $^{48}$Ca extracted from Ref.~\cite{vries} is shown.}
\label{fig5}
\end{figure}
Fig. 4 shows the neutron density of the isotopic chains $^{12-24}$O,  $^{34-70}$Ca, $^{48-98}$Ni, $^{80-150}$Zr, $^{98-176}$Sn and $^{178-262}$Pb as a function of radius where many new neutron bubble nuclei are located. The occupancy (occ.) of 2s, 3s and 4s states near the Fermi level is shown  as a function of neutron number N in the insets of the respective panels of Fig. 4. The zero occupancy of s$_{1/2}$ state in the isotopes of O (N $=$ 8$-$14), Ca (N $=$ 14), Ni (N $=$ 40$-$52), Zr (N $=$ 40$-$60), Sn (N $=$ 48$-$60) and Pb (N $=$ 126$-$158) indicate central density depletion which is seen in their respective neutron density plots Fig. 4 ((a)-(f)). Although the Fermi level is far from the unoccupied 3s state in Ni and Zr isotopes but it appears to influence the central depletion. Interestingly, the mirror nuclei ($^{22}_{8}$O$_{14}$ and $^{22}_{14}$Si$_{8}$) and ($^{34}_{20}$Ca$_{14}$, $^{34}_{14}$Si$_{20}$) (Fig. 4 and Fig. 2) show strong neutron and proton bubbles respectively. $^{80}$Ni$_{52}$, $^{92}$Zr$_{52}$, $^{106}$Sn$_{56}$ and $^{240}$Pb$_{158}$ show significant density depletion. However, the unoccupancy in  2s-state shows stronger density depletion than that in 3s- and 4s-states.\par

\begin{figure}[htb]
\centering
\includegraphics[width=0.48\textwidth]{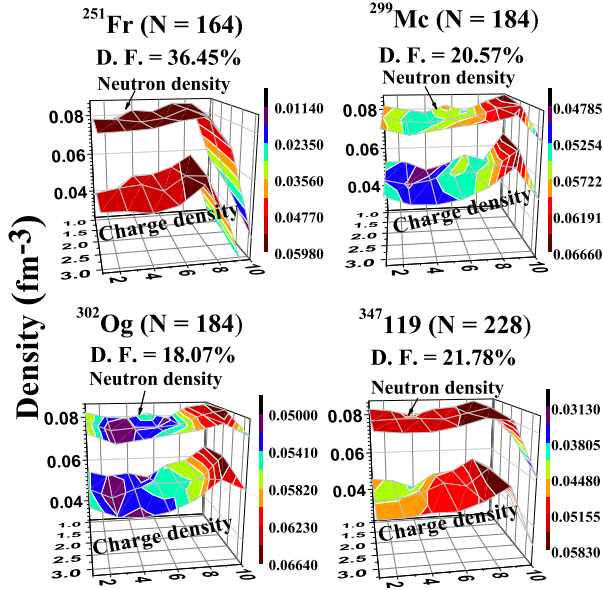}
\caption{(Colour online) Variation of charge density and neuron density is shown for superheavy nuclei $^{251}$Fr, $^{299}$Mc, $^{302}$Og and $^{347}$119.}
\label{fig6}
\end{figure}

The nuclear charge form factor which is a useful physical observable of central depletion, is a measurable quantity through the elastic electron-nucleus scattering experiments~\cite{hofstadter,forest,donnelly}. Form factor for $^{46}$Ar, $^{34}$Si, $^{48}$Ca and $^{36}$S  is displayed in Fig. 5 along with the experimental form factor of $^{48}$Ca~\cite{vries} which shows good agreement. Our computed charge density form factor \cite{duguet,khan} for bubble and non-bubble nuclei differ in their angular distribution, which may be useful for the future experiments to identify proton bubbles by charge density measurements. Upcoming projects like SCRIT~\cite{suda,suda1}, which are expected to provide data to identify proton bubbles by charge density measurements, are still far from reach as far as our predicted nuclei $^{22}$Si, $^{58}$Ar, $^{56}$S, $^{184}$Ce, $^{34}$Ca, $^{80}$Ni, $^{240}$Pb are concerned. However, our predicted bubble nuclei $^{34}$Si, $^{46}$Ar, $^{22}$O, are in principle possible to do with the slow RI beams for use in SCRIT, which may be produced by a projectile fragmentation reaction of $^{70}$Zn beam, and $^{106}$Sn may be produced from $^{124}$Xe beam. RI beams are slowed down by a gas-catcher system to be used in SCRIT. $^{92}$Zr is easy to do even without SCRIT system which appears to be a potential bubble candidate due to its easy experimental accessibility.\par

\begin{figure}[htb]
\centering
\includegraphics[width=0.5\textwidth]{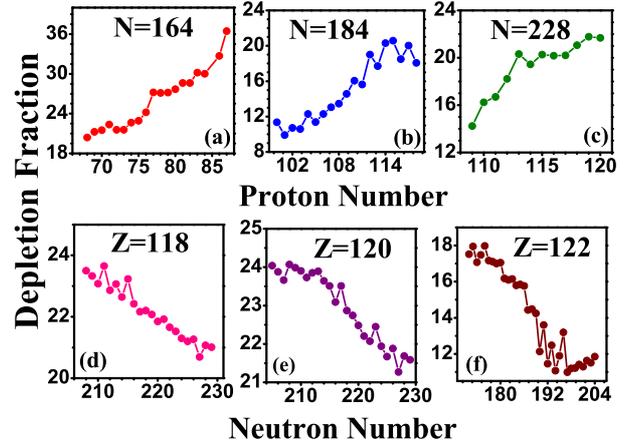}
\caption{(Colour online) Variation of DF vs. Z for N $=$ (a) 164, (b) 184 and (c) 228 isotones. DF vs. N  for Z $=$ (d) 118, (e) 120 and (f) 122 isotopes.}
\label{fig7}
\end{figure}

To search for the bubble like structures in superheavy isotones with neutron numbers N = 164, 184, 228, which have been recently shown to be magic by us~\cite{GAUPRC} and Refs.~\cite{zhang2,adamian,biswal}, we use RMF by including deformation with axially deformed shapes~\cite{gambhir,singh,saxena}. The charge density profiles of $^{251}$Fr, $^{250}$Rn, $^{298}$114, $^{299}$115, $^{301}$117, $^{341}$113 and $^{347}$119 indicate bubble structure with significant central density depletion.
The bubble effect and depletion fraction (DF) of $^{251}$Fr, $^{299}$Mc and $^{347}$119 (shown in Fig. 6) are found much stronger as compared to that of $^{302}$Og (Z $=$ 118) reported in Ref. ~\cite{schuetrumpf}. $^{251}$Fr (Z $=$ 87, N $=$ 164) shows the highest DF that makes it the strongest bubble in the heavy systems seen so far. The depletion fraction (DF) of N $=$ 164, 184, 228 isotones shown in Fig. 7(a),(b),(c) increases with Z due to the increasing Coulomb repulsion. But the DF of Z $=$ 118, 120 and 122 isotopes (plotted in Fig. 7(d), (e), (f)) decreases as neutron number increases indicating the role of isospin and the interplay between the Coulomb and nn-interaction on the bubble effect. Keeping Z or Coulomb interaction fixed, variation of DF is related to isospin and is decreasing as a function of N (i.e. asymmetry (N-Z)/2 parameter). Although the present study shows the influence of Coulomb and asymmetry energy on bubble effect in superheavy systems but some influence of shell effects may also be present as suggested by Ref.$~$\cite{bender} which needs investigation.\par

In Table I, we show  depletion fraction (DF) calculated by using TMA~\cite{singh}, NLSH~\cite{ring3-nlsh}, PK1~\cite{pk1} and NL3*~\cite{nl3star} parameters and compare with various other works~\cite{khan,grasso1,li,schuetrumpf,wu,sksingh,TAN,nakada,yao,thesis,mksharma}. Good agreement between our calculated DF and the other theoretical works validates our predicted known bubble candidates as well as the new potential candidates which are expected to be useful and of interest to future theoretical and experimental studies. \par
\begin{figure}[htb]
\centering
\includegraphics[width=0.40\textwidth]{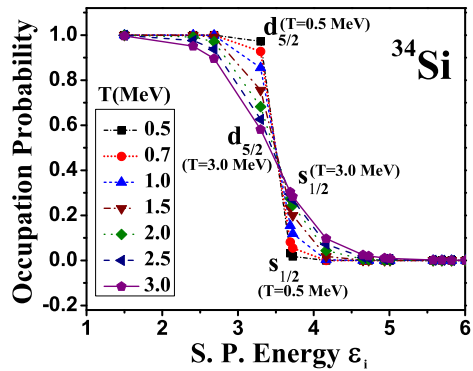}
\caption{(Colour online) Occupation probability as a function of single particle energies ($\epsilon$) at various temperatures T $=$ 0.5$-$3.0 MeV using statistical theory.}
\label{fig8}
\end{figure}

Fig. 8 shows the occupation probability (n$_i$) as a function of single particle energies at various temperatures T $=$ 0.5$-$3.0 MeV using the Statistical theory of hot nuclei~\cite{MAPLB,MAPRC} in the most prominent bubble nucleus $^{34}$Si. As T increases, the occupancy (n$_i$) in 2s$_{1/2}$ state increases from almost 0 to a much higher value which shows the anti bubble effect of temperature. Increasing T washes away the shell effects that leads to changes in the deformation and shape towards sphericity with zero deformation at certain critical temperature (T)~\cite{MAPLB,MAPRC69}. However, since the deformation has a quenching effect on the bubble structure, increasing T may have impact on the deformation as well as the bubble structure. The effect of correlations like pairing correlations ~\cite{grasso,yao1} and shape fluctuations ~\cite{yao1} alter the occupancy but cannot sufficiently quench the bubble structure at zero temperature (T = 0) ~\cite{khan}. In our preliminarily calculations we find that the depletion factor decreases with increasing T and completely vanishes at T = 4 MeV, which is in concise with the recent work~\cite{TAN}. These results along with the detailed study on the impact of temperature on depletion factor would be presented soon in our upcoming work.\par

\begin{table*}
\caption{Depletion fraction DF for our predicted stronger bubble nuclei using RMF parameters TMA~\cite{singh}, NLSH~\cite{ring3-nlsh}, PK1~\cite{pk1} and NL3*~\cite{nl3star}. DF calculated using (a) RMF with FSUGold~\cite{grasso1}, (b) RMF with DDME2~\cite{grasso1}, (c) beyond mean-field calculations with relativistic PCPK1~\cite{yao}, HF method with (d) D1S, (e) D1N and (f) density dependent M3Y-P4 parameters~\cite{thesis}, (g) microscopic non-relativistic HF method: HF(SEI-I)~\cite{mksharma}, (h) Skyrme Hartree-Fock (SHF) using SkI4~\cite{sksingh} and finite temperature HF using MSk3~\cite{TAN} and data extracted from (i) RHFB with PKO3~\cite{li}, (j) RHFB with PKA1~\cite{li}, (k) RMF with PC-PK1~\cite{wu}, (l) HF with M3Y-P7~\cite{nakada}, (m) HF approach with SkI5~\cite{khan} and (n) nuclear density functional theory with Skyrme functionals~\cite{schuetrumpf}}.
\vspace{1.0cm}
\centering
\resizebox{1.0\textwidth}{!}{%
\textbf{
\begin{tabular}{|c|c|c|c|c|c|c|c|c|c|c|c|c|c|}
\hline
\multicolumn{1}{|c|}{N}&\multicolumn{1}{|c|}{Bubble}& \multicolumn{5}{|c|}{Depletion Fraction DF\%}&\multicolumn{1}{|c|}{Z}&\multicolumn{1}{|c|}{Bubble}&
\multicolumn{5}{|c|}{Depletion Fraction DF\%}\\
\cline{3-7}\cline{10-14}
\multicolumn{1}{|c|}{}&\multicolumn{1}{|c|}{Nucleus}& \multicolumn{1}{|c|}{TMA}&\multicolumn{1}{|c|}{NLSH}&\multicolumn{1}{|c|}{PK1}&\multicolumn{1}{|c|}{NL3*}&
\multicolumn{1}{|c|}{Other Theories}&\multicolumn{1}{|c|}{}&\multicolumn{1}{|c|}{Nucleus}& \multicolumn{1}{|c|}{TMA}&\multicolumn{1}{|c|}{NLSH}&
\multicolumn{1}{|c|}{PK1}&\multicolumn{1}{|c|}{NL3*}&\multicolumn{1}{|c|}{Other Theories}\\
\hline
8&$^{22}$Si&14.38&15.30&10.10&7.45&&8&$^{22}$O&31.37&32.64&26.91&17.16&34$^a$, 29$^b$, 27$^d$,\\
&&&&&&&&&&&&&22$^e$, 40$^f$, 11$^{i}$\\
\hline
20&$^{34}$Si&25.23&24.91&23.63&20.10&42$^a$, 36$^b$, 27$^c$, 50$^d$,&20&$^{34}$Ca&40.61&40.32&39.20&34.51&30$^i$ \\
&&&&&& 60$^e$, 37$^f$, 37$^g$, 24$^i$&&&&&&&\\
\hline
28&$^{46}$Ar&42.00&37.67&33.11&20.30&50$^f$, 51$^g$, 16$^i$,&28&$^{80}$Ni&14.22&14.33&14.22&14.18&\\
&&&&&&33$^k$,54$^l$,48$^m$&&&&&&&\\
\hline
40&$^{56}$S &29.63&31.13&26.32&14.58&        &40&$^{92}$Zr&13.38&19.45&13.84&15.12&\\
\cline{8-14}
&$^{58}$Ar&26.03&23.50&18.66&3.79&          &50&$^{106}$Sn&20.42&20.33&20.46&20.36&\\
\hline
126&$^{184}$Ce&14.39&14.03&14.40&14.47& &82&$^{240}$Pb&18.64&19.34&14.70&15.12&\\
\hline
\multicolumn{14}{|c|}{Superheavy Nuclei}\\
\hline
\multicolumn{1}{|c|}{N}&\multicolumn{1}{|c|}{Bubble}& \multicolumn{5}{|c|}{Depletion Fraction DF\%}&\multicolumn{1}{|c|}{N}&\multicolumn{1}{|c|}{Bubble}&
\multicolumn{5}{|c|}{Depletion Fraction DF\%}\\
\cline{3-7}\cline{10-14}
\multicolumn{1}{|c|}{}&\multicolumn{1}{|c|}{Nucleus}& \multicolumn{1}{|c|}{TMA}&\multicolumn{1}{|c|}{NLSH}&\multicolumn{1}{|c|}{PK1}&\multicolumn{1}{|c|}{NL3*}&
\multicolumn{1}{|c|}{Other Theories}&\multicolumn{1}{|c|}{}&\multicolumn{1}{|c|}{Nucleus}& \multicolumn{1}{|c|}{TMA}&\multicolumn{1}{|c|}{NLSH}&
\multicolumn{1}{|c|}{PK1}&\multicolumn{1}{|c|}{NL3*}&\multicolumn{1}{|c|}{Other Theories}\\
\hline
164 &$^{251}$Fr&36.45&34.16&23.59&22.50&&172 &$^{292}$120&17.77&15.89&15.96&15.36&32$^h$, 41$^j$\\
\hline
184   &$^{299}$Mc&20.57&21.23&21.39&17.93&&228&$^{341}$113&20.31&17.31&21.17&19.24&\\
\cline{2-7}\cline{9-14}
&$^{302}$Og&18.07&18.82&18.01&15.62&8$^n$&&$^{347}$119&21.78&21.22&22.38&23.04&\\
\hline
\end{tabular}}}
\label{table1}
\end{table*}

To conclude, the bubble structure in isotopic and isotonic chains of the conventional magic proton (neutron) number P (N)= 8, 20, 28, 40, 50, 82 and N=126, and  recently speculated  magic N $=$ 164, 184, 228 superheavy isotones, is investigated systematically. A complete range of new bubble nuclei is identified in all the mass regions. We employ RMF+BCS approach to study the charge and neutron density profiles, occupation probability and the single particle spectra. This study shows that the central depletion due to unoccupancy in s-orbit is an outcome of shell structure for all the nuclei upto Z $=$ 82. The unoccupied (2s, 3s, 4s) states lead to the proton (neutron) bubble like structure in N (Z) $=$ 8, 20 isotones (isotopes) and neutron bubble in Ni, Zr and Sn and Pb isotopes. The inversion of proton states (2s$_{1/2}$ and 1d$_{3/2}$) and (3s$_{1/2}$ and 1h$_{11/2}$) results in proton bubble nuclei in N $=$ 28, 40 and 126 isotones. Many new superheavy bubble nuclei are traced. The depletion fraction (DF) of magic isotones increases with increasing Z (Coulomb repulsion) and that of the superheavy isotopes decreases with increasing isospin which indicates that the bubble effect is driven by the isospin ((N-Z)/2) and the interplay between the Coulomb and nuclear strong forces. DF calculated by various RMF parameters shows consistency with the other works which shows the validity and usefulness of the RMF+BCS approach for describing bubble nuclei over such a wide range of masses. Charge density form factor for bubble and non-bubble nuclei are found to differ in their angular distribution. To identify the proton bubbles by charge density measurements is still out of reach of the current and near future experimental facilities. However, our theoretical conjectures may be useful for future experiments. Temperature induced effects on bubble nuclei are studied using the statistical theory which indicates the anti bubble effect of temperature. But, variation of bubble effect with T needs more rigorous investigation which would be reported in our subsequent works.  However, the experimental and other theoretical data for our predicted bubble nuclei are anxiously awaited. \par
We thank Prof. H. L. Yadav for his guidance and support. G.S. and M.A. acknowledge the support provided by SERB (DST), Govt. of India under YSS/2015/000952 and WOS-A scheme respectively. We are specially thankful to Prof. M. Wakasugi, Team Leader, SCRIT Team for sharing experimental inputs.


\begin{thebibliography}{99}
\bibitem{WILSON} H.A. Wilson, Phys. Rev. 69 (1946) 538.
\bibitem{nature} A. Mutschler {et al.}, Nature Physics 13 (2017) 152.
\bibitem{duguet} T. Duguet, V. Soma, S. Lecluse, C. Barbieri, and P. Navratil, Phys. Rev. C 95 (2017) 034319.
\bibitem{todd} B.G. Todd-Rudel, J. Piekarewicz, and P.D. Cottle, Phys. Rev. C 69 (2004) 021301(R).
\bibitem{grasso} M. Grasso, Z. Ma, E. Khan, J. Margueron, and Nguyen Van Giai, Phys. Rev. C  76 (2007) 044319.
\bibitem{khan} E. Khan, M. Grasso, J. Margueron, and N. Van Giai, Nucl. Phys. A 800 (2008) 37.
\bibitem{wang} Y.Z. Wang, J.Z. Gu, X.Z. Zhang, and J.M. Dong, Chin. Phys. Lett. 28 (2011) 102101.
\bibitem{wang1} Y.Z. Wang, J.Z. Gu, X.Z. Zhang, and J.M. Dong, Phys. Rev. C 84 (2011) 044333.
\bibitem{grasso1} M. Grasso {et al.}, Phys. Rev. C 79 (2009) 034318.
\bibitem{yao1} Yao, Jiang-Ming {et al.}, Phys. Rev. C 86 (2012) 014310.
\bibitem{li} J.J. Li, W.H. Long, J.L. Song, and Q. Zhao,  Phys. Rev. C 93 (2016) 054312.
\bibitem{schuetrumpf} B. Schuetrumpf, W. Nazarewicz, and P.-G. Reinhard, Phys. Rev. C 96 (2017) 024306 .
\bibitem{wu} X.Y. Wu, J.M. Yao and Z.P. Li, Phys. Rev. C  89 (2014) 017304.
\bibitem{sobi} A. Sobiczewski and K. Pomorski, Prog. Part. and Nucl. Phys. 58 (2007) 292.
\bibitem{decharge} J. Decharge, J.-F. Berger, K. Dietrich, and M.S. Weiss, Phys. Lett. B 451 (1999) 275.
\bibitem{sksingh} S.K. Singh, M. Ikram, and S.K. Patra, Int. J. Mod. Phys. E 22 (2012) 135001.
\bibitem{ikram} M. Ikram, S.K. Singh, A.A. Usmani, and S.K. Patra, Int. J. Mod. Phys. E 23 (2014) 1450052.
\bibitem{bender} M. Bender and P.-H. Heenen, J. Phys. Conf. Ser. 420 (2013) 012002.
\bibitem{TAN} L. Tan Phuc, N. Quang Hung, and N. Dinh Dang, Phys. Rev. C 97 (2018) 024331 .
\bibitem{MAPLB} Mamta Aggarwal, Phys. Lett. B 693 (2010) 489.
\bibitem{MAPRC} Mamta Aggarwal, Phys. Rev. C 89 (2014) 024325.
\bibitem{saxena} G. Saxena, M. Kumawat, M. Kaushik, S.K. Jain, and Mamta Aggarwal, Phys. Lett. B 775 (2017) 126.
\bibitem{saxena1} G. Saxena, M. Kumawat, M. Kaushik, U.K. Singh, S.K Jain, S. Somorendro Singh, and Mamta Aggarwal, Int. J. Mod. Phys. E 26 (2017) 1750072.
\bibitem{walecka} B.D. Serot and J.D. Walecka, Adv. Nucl. Phys. 16 (1986) 1.
\bibitem{gambhir}  P. Ring, Y.K. Gambhir, and G.A. Lalazissis, Comput. Phys. Commun. 105 (1997) 77 and reference therein.
\bibitem{yadav1}  H.L. Yadav, S.Sugimoto, and H. Toki, Mod. Phys. Lett. A 17 (2002) 2523.
\bibitem{singh} D. Singh, G. Saxena, M. Kaushik, H.L. Yadav, and H. Toki, Int. Jour. Mod. Phys. E 21 (2012) 1250076.
\bibitem{decharge1} J. Decharg´e, J.-F. Berger, M. Girod, and K. Dietrich, Nucl. Phys. A 716 (2003) 55.
\bibitem{bender1} M. Bender, K. Rutz, P.-G. Reinhard, J. A. Maruhn, and W. Greiner, Phys. Rev. C 60 (1999) 034304.
\bibitem{yao} J.M. Yao, H. Mei, and Z.P. Li., Phys. Lett. B 723 (2013) 459.
\bibitem{nakada} H. Nakada, K. Sugiura, and J. Margueron, Phys. Rev. C 87 (2013) 067305.
\bibitem{suda} T. Suda and M. Wakasugi, Prog. Part. Nucl. Phys. 55 (2005) 417.
\bibitem{suda1} T. Suda et al., Phys. Rev. Lett. 102 (2009) 102501.
\bibitem{vries} H. De Vries, C.W. De Jager, and C. De Vries, Atomic Data and Nuclear Data Tables 36 (1987) 495536.
\bibitem{chu} Y. Chu, Z. Ren, Z. Wang, and T. Dong, Phys. Rev. C 82 (2010) 024320.
\bibitem{wang2} Y.Z. Wang et al., Phys. Rev. C  91 (2015) 017302.
\bibitem{hofstadter} R. Hofstadter, Rev. Mod. Phys. 28 (1956) 214.
\bibitem{forest} T. de Forest Jr. and J.D. Walecka, Adv. Phys. 15 (1966) 1.
\bibitem{donnelly} T.W. Donnelly and J.D. Walecka, Annu. Rev. Nucl. Part. Sci. 25 (1975) 329.
\bibitem{GAUPRC} G. Saxena, U.K. Singh, M. Kumawat, M. Kaushik, S.K. Jain, and Mamta Aggarwal, Eur. Phys. J. A (2018) (Communicated).
\bibitem{zhang2} W. Zhang, J. Meng, S.Q. Zhang, L.S. Geng, and H. Toki, Nucl. Phys. A 753 (2005) 106.
\bibitem{adamian} G.G. Adamian, N.V. Antonenko, and V.V. Sargsyan, Phys. Rev. C 79 (2009) 054608.
\bibitem{biswal} S.K. Biswal, M. Bhuyan, S.K. Singh, and S.K. Patra, Int. J. Mod. Phys. E 23 (2014) 1450017 .
\bibitem{ring3-nlsh} G.A. Lalazissis, D. Vretenar, and P. Ring, Phys. Rev. C 63 (2001) 034305.
\bibitem{pk1} W. Long, J. Meng, N. Van Giai, and Shan-Gui Zhou, Phys. Rev. C 69 (2004) 034319.
\bibitem{nl3star} G.A. Lalazissis et al., Phys. Lett. B 671 (2009) 36.
\bibitem{thesis} H.S. Than, Ph.D. thesis, Institute of Nuclear Physics of Orsay (2010), www.iaea.org/inis/collection\\/NCLCollectionStore/Public/42/029/42029328.pdf
\bibitem{mksharma} Mahesh K. Sharma, R.N. Panda, Manoj K. Sharma and S.K. Patra, Chinese Physics C 39 (2015) 064102.
\bibitem{MAPRC69} Mamta Aggarwal, Phys. Rev. C 69 (2004) 034602.
\end{thebibliography}
\end{document}